\documentclass[aps,pra,amsfonts,amssymb,epsfig,twocolumn,groupedaddress,showpacs
]{revtex4}
\usepackage{graphics}
\usepackage{epsfig}
\usepackage{amsmath}
\begin{document}
\title{A dynamical model for quantum memory channels}
\author{Vittorio Giovannetti}
\affiliation{NEST-INFM \& Scuola Normale Superiore, I-56126 Pisa, Italy.}
\date{\today}
\begin{abstract}
A dynamical model for  quantum channel is introduced which
allows one to pass continuously from the memoryless case to the case in which 
memory effects are present.
The quantum and classical communication rates of the model are defined 
and explicit expression are provided in some limiting case.
In this context we introduce noise attenuation strategies where part of the
signals are sacrificed to modify the channel environment.
The case of qubit channel with phase damping noise is analyzed in details. 
\end{abstract}
\pacs{03.67.Hk, 03.65.Ud, 89.70.+c}
\maketitle
\section{Introduction}\label{SEC1}
In memoryless quantum channels successive signals (channel uses)
are affected by independent, uniform sources of noise~\cite{CHUANG,SHOR,HSW,SETH,BEN}.
On the other hand, memory channels are characterized by the presence of correlated 
source of noise where each channel use is directly or indirectly affected by the previous ones.
Preliminary results in the study of such systems
has been obtained
in Ref.~\cite{MEMO} where it was pointed out that entangled codes
can be useful in achieving optimal channel performances. Subsequently some of these
results have been generalized to the continuous variable case in Refs.~\cite{GAUS,CERF}, while
a systematic analysis of the problem has been proposed in Refs.~\cite{BOWEN,KRETS}.
In this paper we present a ``dynamical'' model for studying memory effects in quantum communication where 
the noise correlations are derived from
the interactions between  the transmitted signals and 
the channel environment.
By varying the time intervals at which signals are produced by the sender of the message,
the model simulates different communication scenarios.
Memoryless configurations for instance
are recovered as a limiting case in which the
signals are transmitted at 
a frequency much lower than the inverse of the characteristic time of the
channel environment relaxation.
In this context we introduce also {\em noise attenuation} protocols where the sender 
alternates
sequences of carrying-messages signals with sequences of signals 
which are employed to modify the environment response but which do not carry
any messages to the receiver.
Since timescales are fundamental in our model, we characterize its 
efficiency by introducing 
the {\em transmission rates} of the communication line.
These are dimensional quantities (of dimension equal to an inverse time) 
which measure the maximal number of qubits or bits
 of information that can be transferred  reliably (i.e. with unit fidelity) through the
channel {\em per unit of transmitting time}.
Transmission rates are peculiar of our model 
as previous works~\cite{BOWEN,GAUS,MEMO,KRETS,CERF} 
were concerned in characterizing memory channels 
in terms of information capacities, i.e. the maximum number of 
qubits (or bits) that can be reliably transferred through the
channel {\em per channel uses}. 
These figures of merit (i.e. rates and capacities) are in general distinct, but are proportional
to each other 
when the sender of the message encodes her/his messages in regular sequence of signals 
(see Sec.~\ref{s:updown}).

In Sec.~\ref{SEC2} we introduce the channel model by
focusing on the physical assumption which underline its 
definition.
In Sec.~\ref{s:memory} 
we discuss the memory effects present 
in the system and we introduce the noise attenuation protocols.
In Sec.~\ref{SEC3} and 
Sec.~\ref{s:ratechannel} we define the transmission rates of the channel and
we compute their values in some extremal case.
Finally in Sec.~\ref{SEC4} an 
 example 
of a dephasing qubit channel with memory is discussed.
 
\section{The model}\label{SEC2}

Consider a communication line where
messages are encoded
into some internal
degree of freedom (e.g. polarization, spin etc.) of a 
collection of identical physical objects C$_1$, C$_2$, $\cdots$  
which propagate 
through the medium E that
separates the sender (say Alice) from the receiver (Bob).
The C$_j$ are the information  carriers 
of the system: they are locally produced by Alice  and organized 
in a time-ordered sequence ${s}= \{\tau_1, \tau_2, \cdots \}$
with $\tau_j>0$ being the time interval between 
the instants $t_{j+1}$ and $t_{j}$ at which C$_{j+1}$ and C$_j$ 
enter E respectively. 
We will assume 
the effective
 transit time ${\cal  T}_{tr}$ it takes for the carriers for reaching
Bob 
to be  constant and shorter than the intervals $\tau_j$ at which they
are injected into the medium ({\em fast propagation condition}). 
The first condition 
guarantees that the time-ordering of 
${s}$ 
is  preserved in the propagation
(i.e. Bob will receive the 
$({j+1})$-th carrier only after a time $\tau_j$ 
from the arrival of the $j$-th carrier).
The second condition instead
guarantees
that E interacts only with one carrier at a time.
Therefore, if $R$ is 
the density matrix 
of the carriers at Alice location, after a time 
${\cal T}_{tr}$ Bob will receive the state 
\begin{eqnarray}
R^\prime =
\mbox{Tr}_E  \; \big\{ W \; (R \otimes \rho_0 )
\; W^\dag  \big\} \label{prima}\;,
\end{eqnarray} 
where $\rho_0$ is the initial state of E, and 
where 
\begin{eqnarray}
W = \cdots 
V_{j} U_{j}\; \cdots \; 
V_2 U_{2} \; V_1 U_{1}
\;, \label{terza}
\end{eqnarray}
is  the 
unitary operator which describes the 
coupling between the internal degree of freedom of  the carriers
and~E.
In Eq.~(\ref{terza}) the terms
 \begin{eqnarray}
U_{j}  \equiv 
 T \exp\left\{ -\frac{i}{\hbar} \int_{t_{j}}^{t_{j}+{\cal T}_{tr}} dt \; \left[ 
H_{C_j E} (t) + H_E \right] \right\}
\;, \label{seconda3}
\end{eqnarray} 
describe the interaction 
between C$_j$ and E
(here
$H_{C_j E}(t)$ is the effective
time dependent Hamiltonian that couples 
C$_j$ and E, while  $H_E$ is the
free Hamiltonian of the medium).
Working in a {\em strong coupling regime} we will neglect the contribution of $H_E$
in Eq.~(\ref{seconda3})  
and we will assume the $U_j$ to be uniform with respect to the label~$j$.
On the other hand,
the terms $V_j$ of Eq.~(\ref{terza}) describe
the free evolution of E in the time interval between  the instant $t_j+{\cal T}_{tr}$
when 
C$_j$ leaves the environment 
and the instant  $t_{j+1}$  when C$_{j+1}$ enters it, i.e.
\begin{eqnarray}
V_{j} \equiv \exp\left\{ -\frac{i}{\hbar} H_E (\tau_j- {\cal T}_{tr}) \right\} \simeq
\exp\left\{ -\frac{i}{\hbar} H_E \tau_j \right\}
\;. \label{seconda4}
\end{eqnarray} 
\begin{figure}[t]
\begin{center}
\epsfxsize=.6\hsize\leavevmode\epsffile{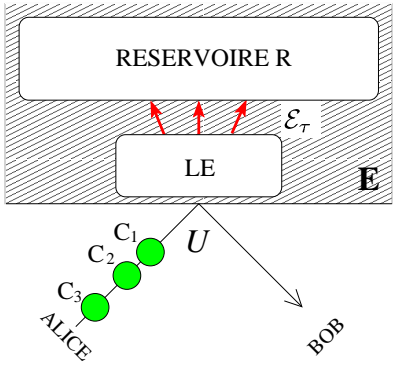}
\end{center}
\caption{Schematic of the communication scenario.
Alice encodes her messages in the internal degree of freedom of the
carriers  C$_1$, C$_2$, $\cdots$,  which propagates in a time-ordered sequence
toward Bob. 
The carriers interact one at a time with the local environment LE,
 while  LE undergoes a dissipative evolution through its interaction with
the reservoir R.}
\label{f:figu0}
\end{figure}
In the following we identify two distinct components of the
medium E: a finite dimensional Local Environment (LE) component
which is directly coupled
with the carriers through the $U_{j}$, and a huge Reservoir (R) 
component
which is coupled with LE but not with the carriers (see Fig.~\ref{f:figu0}). 
The free evolution~(\ref{seconda4}) is supposed  to
induce a dissipative dynamics  which
transforms any initial states of LE into a stationary 
configuration $\sigma_0$, with 
$\tau_E$ being the characteristic time of the process.
This is equivalent~\cite{PETRU} to introducing  
a one-parameter family 
${\cal F}\equiv \{{\cal E}_\tau\}_{\tau\geqslant 0}$ 
of  Completely Positive, Trace preserving
(CPT) which, 
given $\sigma$ the initial state of LE at some time $t_0$, 
represents its evolution at time $t_0+\tau$ with the density matrix
${\cal E}_\tau(\sigma)$.
In this formalism ${\cal E}_0$ coincides with 
identity map on ${\cal H}_{LE}$. On the other hand
the stationary state $\sigma_0$ of LE
is defined by the property
\begin{eqnarray}
{\cal E}_\tau(\sigma_0) &=& \sigma_0 \quad 
\mbox{for all $\tau\geqslant 0$}\;,
\label{relax}
\end{eqnarray}
while the characteristic time $\tau_E$ by the property
\begin{eqnarray}
{\cal E}_{\tau\geqslant \tau_E} 
(\Theta) = \sigma_0\;  \mbox{Tr}\;  \Theta \label{mappaTAUE}\;,
\end{eqnarray}
for all bounded operator $\Theta$ of ${\cal H}_{LE}$.
An 
example of $\cal F$ satisfying the above conditions will be 
presented in Sec.~\ref{SEC4}.

\begin{figure}[t]
\begin{center}
\epsfxsize=.99\hsize\leavevmode\epsffile{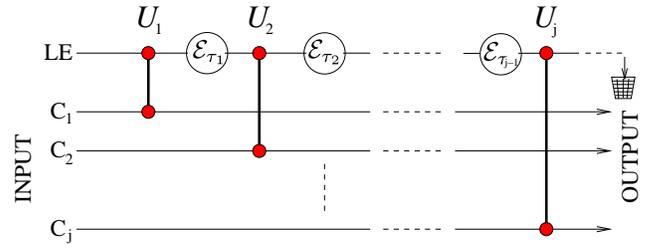}
\end{center}
\caption{Circuit representation of Eq.~(\ref{mappaENNE}).
The local environment LE interacts through the unitary couplings $U_j$ 
(represented by the small red circles in the figure) with 
one carriers at a time. Between two consecutive interactions with the carriers instead LE 
undergoes the dissipative evolution described by the transformations ${\cal E}_{\tau_j}$
(open circles).}
\label{f:figu1}
\end{figure}

Under the above approximations Eq.~(\ref{prima}) provides
a {\em bouncing ball} description of the carrier-environment 
interactions  where
the carriers-balls move toward the LE-wall according to the time-ordered
sequence ${s} = \{\tau_1, \tau_2,
\cdots \}$ chosen by the ``pitcher'' Alice and 
``hit'' instantaneously the local environment LE one at a time
(see Fig.~\ref{f:figu0}). 
The resulting transformation is 
a time ordered product of interactions  $U_j$ 
and relaxation processes  ${\cal E}_{\tau_j}$ (see Fig.~\ref{f:figu1}).
Assuming LE to be initially in 
the stationary state $\sigma_0$ this gives
\begin{eqnarray}
R^\prime 
&=& 
\mbox{Tr}_{LE}  \; \big\{ \cdots \circ {\cal E}_{\tau_{j}} \circ {\cal U}_j \circ
\cdots
\nonumber \\
&&\qquad \qquad \circ  {\cal E}_{\tau_{2}} \circ {\cal U}_2
 \circ {\cal E}_{\tau_{1}}
\circ {\cal U}_{1} 
\; (R \otimes \sigma_0 ) \big\} 
\;,
\label{mappaENNE}
\end{eqnarray} 
where the partial trace is performed on ${\cal H}_{LE}$, ${\cal U}_j(\cdots)$ 
stands for  the unitary mapping $U_j (\cdots ) U_j^\dag$ on 
${\cal H}_{C_j}\otimes{\cal H}_{LE}$, and ``$\circ$'' indicates  the composition
of super-operators.
It is important to note that in our model each sequence
${s}=\{\tau_1,\tau_2, \cdots\}$ is characterized by
a distinct input-output relation~(\ref{mappaENNE}).
\section{Memory effects}\label{s:memory}

Here we give an overview of the memory effects 
which are accounted for by the model introduced in Sec.~\ref{SEC2}.

Because of the time ordering of Eq.~(\ref{mappaENNE})
the output state of a carrier might depend on the input state of the carriers
which precedes it in $s$ but it is always
independent from the input state of the carriers  
which follows it in the sequence.
As  a matter of fact Eq.~(\ref{mappaENNE})  
closely resembles the
memory channels analyzed by Kretschmann and Werner in Ref.~\cite{KRETS}.
To make this more explicit we 
rewrite this equation in terms of 
of the discrete family of CPT  maps $\{\Phi_{s}^{(n)}\}_n$
where
\begin{eqnarray}
\Phi_{s}^{(n)} (R)   
\nonumber 
&\equiv&\mbox{Tr}_{LE}  \; \big\{ {\cal U}_{n} 
\circ {\cal E}_{\tau_{n-1}}
\circ {\cal U}_{n-1}\\ 
&& \qquad  \circ 
\cdots  
\circ {\cal E}_{\tau_{1}}
\circ {\cal U}_{1} 
\; (R \otimes \sigma_0 ) \big\} 
\label{APmappaENNE}
 \;,
\end{eqnarray}
is the output state~(\ref{mappaENNE}) corresponding to the
 density matrix $R$ of $\otimes_{j=1}^{n}
{\cal H}_{C_j}$ associated with the first $n$ carriers of the sequence ${s}$ 
(here ${\cal H}_{C_j}$ is the Hilbert
space associated with the internal degree of freedom of the $j$-th carrier).
Therefore the model of Sec.~\ref{SEC1}
originates proper memory effects analogous to those
of Refs.~\cite{BOWEN,KRETS,MEMO,CERF} but avoids the feed-forward
correlations of Ref.~\cite{GAUS}. 
For instance Markovian correlated noise can be recovered by properly choosing 
the transformations ${\cal E}_{\tau_j}$ (see Appendix \ref{appendixA}).

\subsection{Memoryless configuration}\label{SEX}

\begin{figure}[t]
\begin{center}
\epsfxsize=.9\hsize\leavevmode\epsffile{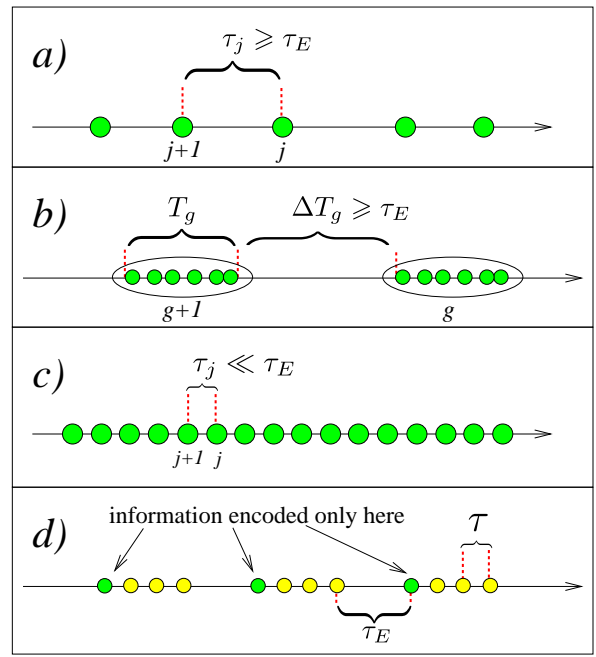}
\end{center}
\caption{Some relevant configurations. Part~{\em a)}: memoryless configuration~(\ref{mappaENNE1}).
The carriers (represented by the green circles) are separated by
time intervals $\tau_j$ which are greater than the dissipation time $\tau_E$ of the local environment.
Part~{\em b)}: generalized memoryless configuration~(\ref{mappaENNE30}).
Here the carriers are divided in groups labeled by the index~$g$. The groups are separated by
time intervals  $\Delta T_g$ which are greater than the dissipation time $\tau_E$.
Part~{\em c)}: perfect memory channel~(\ref{mappaENNE10}). Here
the distance between two consecutive carriers is negligible with respect to $\tau_E$ inhibiting
the relaxation of LE.
Part~{\em d)}: example of a noise attenuation protocol. 
Alice sends uniform sequences of signals composed by~$n$ 
carriers (the B carriers of the protocol represented by yellow circles in the picture) 
which have been prepared in the same input state $\rho_0$  and which
are separated by time intervals $\tau$. These carriers do not convey any message to
Bob and are employed only to ``program'' the environment response.
The information is instead encoded into
the $(n+1)$-th  carrier (the A carriers of the protocol represented by the  green circles). 
The sequence repeats after a time interval $\tau_E$ to
allow LE to return to the stationary configuration.}
\label{f:figu5}
\end{figure}

Assume Alice is producing a sequence ${s}$
with intervals $\tau_j$ greater than or equal to the
characteristic relaxation time $\tau_E$ of the dissipation process $\cal F$-- see part a) of Fig.~\ref{f:figu5}. 
In this case, after each interaction,  the local environment LE has enough time to 
relax into the stationary configuration $\sigma_0$ 
before a new carrier begins interacting with it.
Under this hypothesis Eqs.~(\ref{mappaTAUE}) 
and~(\ref{APmappaENNE}) yield
 \begin{eqnarray}
\Phi_{s}^{(n)} = {\cal N}^{\otimes n} 
\label{mappaENNE1} 
\end{eqnarray} 
where ${\cal N}$ is the CPT map 
which transforms the
density matrices $\rho$ of a single carrier into 
 \begin{eqnarray}
{\cal N}(\rho)  =  \label{mappaENNE2} 
\mbox{Tr}_{LE}  \; \big\{ {\cal U} 
(\rho  \otimes \sigma_0 ) \big\} \;.
\end{eqnarray}
Equation~(\ref{mappaENNE1}) describes a memoryless configuration 
where the noise acts on the C$_j$  independently.
\subsection{Generalized memoryless configuration}
\label{SEX1}
A generalization of~(\ref{mappaENNE1}) is obtained
when the carriers are
organized in identical independent groups of $m$ elements 
each. 
Here it is convenient to express the elements of $s$ 
as $\tau_{g,\ell}$ where $g=1,2,\cdots$ is the group index,
while $\ell\in\{1, \cdots, m\}$ labels the carriers 
within a given group. In this notation the time interval 
\begin{eqnarray}
T_g = \sum_{\ell=1}^{m-1} \tau_{g,\ell}\;,
\label{imp}
\end{eqnarray}
gives the ``length''  of the $g$-th group
while $\Delta T_g=\tau_{g,m}$ is the interval
which separates the last element of the $g$-th group from the first
element of the $(g+1)$-th group.
We do not assume any restrictions on
the time intervals 
$\{\tau_{g,\ell}\}_{\ell = 1, \cdots, m-1}$ 
which separates carriers 
belonging
to the same group
but we require carriers of 
distinct subgroups to be separated by time
intervals larger than $\tau_E$, i.e.
$\Delta T_{g} \geqslant \tau_E$ -- see Fig.~\ref{f:figu5} part~{\em b)}. 
In this case  from Eq.~(\ref{APmappaENNE}) 
follows that the transformation of the carriers of the first
$G$ groups can be expressed as 
\begin{eqnarray}
\Phi_{s}^{(n)}  =  \label{mappaENNE30} 
\otimes_{g=1}^{G} {\cal M}_s^{(g)} \;,
\end{eqnarray} 
where  $n=mG$ and
 \begin{eqnarray}
&&{\cal M}_{s}^{(g)} (\rho)  \equiv 
\label{mappaENNE40}
\\ 
&&\quad \mbox{Tr}_{LE} \big\{ {\cal U}_{g,m} \circ {\cal E}_{\tau_{g,{m-1}}}
\circ \cdots  
\circ {\cal E}_{\tau_{g,1}}
\circ {\cal U}_{g,1} 
(\rho  \otimes \sigma_0 ) \big\} \;,\nonumber
\end{eqnarray}
is the CPT map associated with the $m$ carriers 
C$_{g,1}$, $\cdots$, C$_{g,m}$ of the 
$g$-th group. By comparison with Eq.~(\ref{mappaENNE1}),
 Eq.~(\ref{mappaENNE30}) describes a memoryless channel 
where the groups are the effective information
carriers of the model.
In particular if the sets
 $\{ \tau_{g,\ell}\}_{\ell = 1,\cdots, m}$ are uniform with respect to the
 group label $g$, one has ${\cal M}_s^{(g)}={\cal M}_s^{(g^\prime)}$ for all
$g$ and $g^\prime$ and the transformation~(\ref{mappaENNE30}) 
has once again the standard tensor structure ${\cal M}_s^{\otimes G}$.

\subsection{Perfect memory channel}
\label{PERF}
Consider the case where $\tau_j \ll \tau_E$ for all $j$. 
In this limit the local environment
relaxation process is inhibited by the frequent
interactions with the carriers. Consequently the ${\cal E}_{\tau_j}$ 
are replaced by the identity transformation on ${\cal H}_{LE}$ 
and Eq.~(\ref{APmappaENNE}) yields
\begin{eqnarray}
\Phi_s^{(n)}(R)
\label{mappaENNE10} 
= \mbox{Tr}_{LE}  \; \big\{ {\cal U}_n \circ \cdots  
\circ {\cal U}_2 \circ {\cal U}_{1} 
\; (R \otimes \sigma_0 ) \big\} \;.
\end{eqnarray} 
This expression
describes 
a perfect memory channel~\cite{BOWEN,KRETS} 
where the information transferred from  the 
carriers to the finite 
dimensional local environment LE is not dissipated into
the reservoir R of Fig.~\ref{f:figu0}. These maps are asymptotically equivalent~\cite{KRETS} to
noiseless channel where each carriers can transfer $\log_2 D$
qubits of quantum information reliably
(here $D$ is the dimension of the Hilbert
space ${\cal H}_C$ of a single carrier).

\subsection{Noise attenuation protocols}\label{s:noiseatt}

Here we present a 
communication strategy which explicitly exploits the fact that
in our model the environment is effected by the signaling process.
In this protocols only a subset A of the 
transmitted carriers is used to encode messages to Bob. The remaining  
carriers (subset B) 
are instead employed  for perturbing LE in such a way that
the C$_j$ on which the messages are encoded have a better chance
to reach Bob without being corrupted.
In other words the B carriers are used by the sender as control
parameters to program the environment response to the A carriers.
A simple implementation of a noise attenuation scheme
is shown in Fig.~\ref{f:figu5}~part {\em d)}.
Here the B carriers
are composed by uniform strings of $n$ states $\rho_0$ (represented by 
the yellow circles)
separated by equal time intervals $\tau$.
The information is instead  encoded a single carrier (green circles) and
the whole structure repeats after a relaxation
time $\tau_E$ -- this last assumption is not fundamental but allows us to treat
the input-output relations 
of the A carriers as a memoryless channel of the form~(\ref{mappaENNE1}).
In this configuration the transformation of the A carriers which
comes from solving Eq.~(\ref{mappaENNE}) can be computed as follows.
First we determine the modified state $\sigma_n$ of LE
which arises from the
interactions with
the B carriers. This is accomplished by solving the set of coupled equations
analogous to those of Ref.~\cite{SCARANI},
\begin{eqnarray}
\left\{ 
\begin{array}{l}
\sigma_j^\prime = \mbox{Tr}_C \;\{ {\cal U}(\rho_0 \otimes \sigma_j) \} \;, \\
\sigma_{j+1} = {\cal E}_{\tau} (\sigma_{j}^\prime)   \;,
\end{array} 
\right.
\label{MinputoutputNEW}
\end{eqnarray}
where the trace is performed over the carrier degree of freedom, ${\cal U}$ is
the usual carrier-LE coupling super-operator and $j=0,1, \cdots, n-1$.
The density matrix $\sigma_n$ which results from~(\ref{MinputoutputNEW})
is then used to determine the output state of the A carriers according to
the equation
\begin{eqnarray}
\overline{\cal N}(\rho) 
\equiv  \mbox{Tr}_{LE} \{ {\cal U} (\rho \otimes \sigma_{n}) \} \;.
\label{MinputoutputNEW1}
\end{eqnarray}
The transformation (\ref{MinputoutputNEW1})  is in general different from 
Eq.~(\ref{mappaENNE2}) and depends explicitly on the parameters $n$, $\tau$ 
and $\rho_0$ that are controlled by Alice. The basic idea of a noise attenuation
scheme is to appropriately select such parameters
in order to get a transformed mapping  $\overline{\cal N}$ which is less noisy than the original
mapping $\cal N$.
An example of this effect will
be presented 
in Sec.~\ref{SEC4}.

\section{Transmission rate of a sequence}\label{SEC3}

Timescales play a fundamental role in the model presented in Sec.~\ref{SEC2}. 
Therefore
a proper way to characterize it, is by introducing its quantum and classical 
transmission rates. 
In simple terms these quantities measure, respectively, 
the maximum number of qubits and bits per second that Alice 
can encode into the carriers sequence $s$ 
without compromising the readability
of the transmitted messages. The formal definition of  the rate of the sequence $s$  
is constructed as
 follows.

First of all we introduce the
discrete value function 
$n_s(T)$ which, given the sequence $s$, 
counts the number of carriers which fit~\cite{NOTAZERO}
in the time interval $[0,T[$.
Furthermore, 
for any $\epsilon >0$  and $T >0$ 
we  define 
$q_{s}(\epsilon,T)$ to be 
the dimension --in qubits units-- of the largest Hilbert sub-space of ${\cal H}(T)
\equiv \otimes_{j=1}^{n_s(T)} {\cal H}_{C_j}$ 
which allows for a fidelity of the transmitted
state greater than $1-\epsilon$. This is
\begin{eqnarray}
q_s({\epsilon},T)
=
\max_{d}
 \Big\{ \log_2 d  : \exists {\cal H}_{code} \; \dim {\cal H}_{code}=d , \; \exists \;  
{\cal A}, 
{\cal D}
\nonumber  \\
\forall |\Psi\rangle \in {\cal H}_{code} \;\;
F(\Psi,{\cal D}\circ \Phi_{s}^{(T)}\circ{\cal A})>1-\epsilon \;
\Big\},\quad \; \label{formularate2} 
\end{eqnarray}
where ${\cal H}_{code}$ are Hilbert sub-spaces of  ${\cal H}(T)$, 
${\cal A}$ and ${\cal D}$ are {\em encoding} and {\em decoding} CPT
maps on ${\cal H}(T)$ applied, respecitively, by Alice and Bob to the carriers,
and 
\begin{eqnarray} 
F(\Psi,{\cal D}\circ\Phi_{s}^{(T)}\circ{\cal A} )\equiv \langle \Psi |{\cal D} 
\circ \Phi_{s}^{(T)}\circ{\cal A}(|\Psi\rangle\langle
\Psi|) | \Psi\rangle 
\label{fidelity}\;,
\end{eqnarray}
is the fidelity between the input state $|\Psi\rangle \in {\cal H}_{code}$
and the {\em decoded} output state ${\cal D} 
\circ \Phi_{s}^{(T)}\circ {\cal A} (|\Psi\rangle\langle
\Psi|)$ (for easy of notation $\Phi_{s}^{(T)}$ indicates
the map $\Phi_{s}^{(n_s(T))}$ of Eq.~(\ref{APmappaENNE}) 
that acts on the $n_s(T)$ carriers of $s$ which lie on $[0,T[$).
 The quantum transmission rate $r_q(s)$ of $s$ 
is thus given by the ratio  $q_s({\epsilon},T)/T$ in the 
the limits $\epsilon\rightarrow 0$, $T\rightarrow \infty$
, i.e. \cite{NOTASUP}
\begin{eqnarray}
r_q(s) = \lim_{\epsilon \rightarrow 0} \;
\limsup_{T\rightarrow \infty} \;  
\frac{ 
q_{s}(\epsilon,T)}{T}\;.
\label{formalrate1}\end{eqnarray}
Analogously we define the {\em classical} transmission rate $r_c(s)$ of $s$
by substituting the function
$q_s({\epsilon},T)$ with
the largest number of classical distinguishable messages 
$c_s({\epsilon},T)$ that can be transmitted to
Bob with fidelity greater than $1-\epsilon$, i.e.
\begin{eqnarray}
r_c(s) = \lim_{\epsilon \rightarrow 0} \;
\limsup_{T\rightarrow \infty} \; \frac{c_s({\epsilon},T)}{T}\;,
\label{formalrate10}\end{eqnarray} 
where as in Eq.~(\ref{formularate2}) one has
\begin{eqnarray}
c_s({\epsilon},T)
=
\max_{d}
 \Big\{ \log_2 d  : \exists {\cal H}_{code} \; \dim {\cal H}_{code}=d , \; 
\exists \;  {\cal A}, {\cal D}
\nonumber  \\
\forall k\in\{1, \cdots, d\}   \;\;
F(\Psi_k,{\cal D}\circ \Phi_{s}^{(T)}\circ{\cal A} )>1-\epsilon \;
\Big\}, \; \label{formularate3} 
\end{eqnarray}
with $|\Psi_1\rangle,|\Psi_2\rangle, \cdots, |\Psi_d\rangle$
being an orthonormal basis of ${\cal H}_{code}$.

\subsection{Upper and lower bounds}\label{s:updown}
A simple upper bound for the quantum rate $r_q({s})$ of $s$  
can be derived from Eq.~(\ref{formalrate1}) as
follows,~\cite{NOTALIM}
\begin{eqnarray}
r_q({s}) &=& \lim_{\epsilon \rightarrow 0} \;
\limsup_{T\rightarrow \infty} \; \frac{n_s(T)}{T}  \frac{q_s({\epsilon},T)}{n_s(T)} \nonumber \\
&\leqslant &\Big[ \lim_{\epsilon \rightarrow 0} \;
  \limsup_{T\rightarrow \infty}  \frac{q_s({\epsilon},T)}{n_s(T)}\Big]\;  
\limsup_{T^\prime \rightarrow \infty} \; \frac{n_s(T^\prime)}{T^\prime}\nonumber \\
&=& Q_s/ {\tau}^\prime_s \;, \label{upper1}
\end{eqnarray}
where ${\tau}^\prime_s$   is the {\em minimum
average  first-neighbors distance} among the carriers of ${s}$ defined by
\begin{eqnarray}
1/{\tau}^\prime_s  = \limsup_{T^\prime\rightarrow \infty} \; \frac{n_s(T^\prime)}{T^\prime}=
\lim_{T^\prime \rightarrow \infty} \sup_{t\geqslant T^\prime}  \frac{n_s(t)}{t}\;.
\label{unosutauprimo}
\end{eqnarray}
On the other hand
\begin{eqnarray}
Q_s = \lim_{\epsilon \rightarrow 0} \limsup_{T\rightarrow \infty} \; 
\frac{q_s({\epsilon},T)}{n_s(T)}= \lim_{\epsilon\rightarrow 0} 
\limsup_{n \rightarrow \infty} \frac{q_s({\epsilon},n)}{n}
\label{qdis}
\end{eqnarray}
defines  the quantum capacity~\cite{BARNUM,KRETS,KRETS1,SHOR} associated with  the maps 
$\{\Phi_{s}^{(n)}\}_n$ of Eq.~(\ref{APmappaENNE})
(in this expression  $q_s({\epsilon,n})$ is given by (\ref{formularate2})
with $n_s(T)$ replaced by $n$).

A lower bound for $r_q(s)$ is instead obtained as follows~\cite{NOTALIM}
\begin{eqnarray}
r_q({s}) &=& \lim_{\epsilon \rightarrow 0} \;
\limsup_{T\rightarrow \infty} \; \frac{n_s(T)}{T}  \frac{q_s({\epsilon},T)}{n_s(T)} \nonumber \\
&\geqslant &\Big[ \lim_{\epsilon \rightarrow 0} \;
  \limsup_{T\rightarrow \infty}  \frac{q_s({\epsilon},T)}{n_s(T)}\Big]\;  
\liminf_{T^\prime \rightarrow \infty} \; \frac{n_s(T^\prime)}{T^\prime}\nonumber \\
&=& Q_s/ {\tau}^{\prime \prime}_s \;, \label{lower1}
\end{eqnarray}
where ${\tau}^{\prime\prime}_s\geqslant {\tau}^\prime_s$   is the {\em maximum
 first-neighbors  average distance} among the carriers of ${s}$ defined by
\begin{eqnarray}
1/{\tau}^{\prime\prime}_s  = \liminf_{T^\prime\rightarrow \infty} \; \frac{n_s(T^\prime)}{T^\prime}=
\lim_{T^\prime \rightarrow \infty} \inf_{t\geqslant T^\prime}  \frac{n_s(t)}{t}\;.
\label{unosutausecondo}
\end{eqnarray}
If the sequences $s$ is such that 
$\lim_{T\rightarrow \infty}{n_s(T)}/{T} = 1/\tau_s$ exists, 
one has $\tau^\prime_s=\tau^{\prime\prime}_s=\tau_s$ 
with $\tau_s$ being the average first-neighbors distance among the carriers.
These are the {\em regular} sequences of the model: 
for them Eqs.~(\ref{upper1}) and  (\ref{lower1})
coincide and the transmission rate is proportional to the quantum capacity of the channel, i.e.
\begin{eqnarray}
r_q({s}) 
&=& Q_s/ {\tau}_s \;. \label{ratedis}
\end{eqnarray}
The same analysis can be repeated also for  the classical rate $r_c(s)$ of Eq.~(\ref{formalrate10}).
In particular, in this case,
 Eqs.~(\ref{upper1}), (\ref{lower1}) 
and (\ref{ratedis}) still apply
by replacing $Q_s$ with the classical capacity $C_s$ of the maps $\{\Phi_s^{(n)}\}_n$
defined by 
\begin{eqnarray}
C_s 
= \lim_{\epsilon\rightarrow 0} 
\limsup_{n \rightarrow \infty} \frac{c_s({\epsilon},n)}{n}\;.
\label{cdis}
\end{eqnarray}

\subsection{Some solvable configurations}
\label{s:solvable}

The maximizations implicit in Eqs.~(\ref{qdis}) and (\ref{cdis})  are  in general
difficult to solve. However, following the analysis of 
Refs.~\cite{KRETS,BARNUM} one can bound the capacities
$Q_s$ and $C_s$ by means of the coherent information~\cite{COHEINFO} and of 
the Holevo information~\cite{HOLEVO} of $\Phi_s^{(n)}$, respectively.
In particular we have
\begin{eqnarray}
Q_s &\leqslant& 
\limsup_{N \rightarrow \infty}  \max_{R} 
\; \frac{J (\Phi_{s}^{(N)}, R)}{N}  \;,
 \label{quantum1111}
\end{eqnarray}
where the maximization is performed over all density
matrices $R$ of $N$ carriers and
\begin{eqnarray}
 J(\Phi_{s}^{(N)}, R) \equiv S(\Phi_{s}^{(N)}(R))
-S((\Phi_{s}^{(N)}\otimes {\cal I}_A)(\Psi_R)),
\label{COHERENT}
\end{eqnarray}
is the coherent information~\cite{COHEINFO} of $\Phi_{s}^{(N)}(R)$.
In the above expression $S(R) = -\mbox{Tr}[R \log_2 R]$
is the von Neumann entropy, $\Psi_R$ is a generic purification
of $R$ constructed by adding an ancillary Hilbert space
${\cal H}_A$, and ${\cal I}_A$ is the identical map
on ${\cal H}_A$.
Analogously one has
\begin{eqnarray}
C_s &\leqslant& 
\limsup_{N \rightarrow \infty} 
\; \max_{\cal P}\; \frac{\chi(\Phi_s^{(N)}, {\cal P})}{N}
\label{ccCC}
\end{eqnarray}
where the maximization is performed over all
ensemble ${\cal P} =\{p_k; R_k\}_k$ of $N$ carriers and where
\begin{eqnarray} 
\chi(\Phi_s^{(N)}, {\cal P})&\equiv& 
S(\Phi_s^{(N)}(\sum_{k} p_k R_{k}))
\label{c} \\
&&\qquad -\sum_k p_k
S(\Phi_s^{(N)}(R_k))\;, \nonumber 
\end{eqnarray}
is the  Holevo information~\cite{HOLEVO} associated with 
$\Phi_s^{(N)}$. 
Kretschmann and Werner have identified
a class of maps $\{ \Phi_s^{(n)}\}_n$  (the {\em forgetful} channels~\cite{KRETS})
for which the right-hand side term of~(\ref{quantum1111}) 
and~(\ref{ccCC}) indeed provide the exact value for 
$Q_s$ and $C_s$. 
Here we will focus only on the limiting cases discussed in Sec.~\ref{s:memory}
for which an expression for $Q_s$ and $C_s$ can be derived without the elegant
arguments of Ref.~\cite{KRETS}.
\begin{itemize}
\item[{\bf \em a)}]
The simplest configuration is when
the sequence $s$ is such that $\tau_j \ll \tau_E$ for all $j$. When this happens
the maps $\{\Phi_s^{(n)}\}_n$ describe a perfect memory channel~(\ref{mappaENNE10})
which allows optimal transfer, ensuring $Q_s = C_s =\log_2 D$.
Therefore, according to~(\ref{ratedis}) using regular sequences $s$ with $\tau_j\ll \tau_E$,  
Alice and Bob can achieve
transmission rates equal to
\begin{eqnarray}
r_q({s}) = r_c({s})=\frac{\log_2 D}{ {\tau}_s} \;. \label{ratesperfect}
\end{eqnarray}
\item[{\bf \em b)}]
For memoryless 
configurations~(\ref{mappaENNE1}),
$Q_s$ and $C_s$
coincide, respectively, with the quantum $Q({\cal N})$ and classical $C({\cal N})$ capacity of the
memoryless map $\cal N$ of Eq.~(\ref{mappaENNE2}). On one hand one has~\cite{SETH},
\begin{eqnarray}
{Q({\cal N})} = \lim_{N \rightarrow \infty}  \max_{R} 
\; \frac{J ({\cal N}^{\otimes N}, R)}{N}  \;, 
\label{quantum11}
\end{eqnarray} 
where, as in Eq.~(\ref{quantum1111})
 the maximization is performed over all density
matrices $R$ of $N$ carriers and where $J({\cal N}^{\otimes N}, R)$
is the coherent information~(\ref{COHERENT}) of ${\cal N}^{\otimes N}$.
On the other hand one has~\cite{HSW},
\begin{eqnarray}
C({\cal N}) = \lim_{N \rightarrow \infty} 
 \max_{\cal P}\; \frac{\chi({\cal N}^{\otimes N}, {\cal P})}{N}
\label{cc}
\end{eqnarray}
where the maximization is performed over all
ensemble ${\cal P} =\{p_k; R_k\}_k$ of $N$ carriers and where $\chi({\cal N}^{\otimes N}, {\cal P})$
is the  Holevo information~(\ref{c}) associated with 
${\cal N}^{\otimes N}$. 
Therefore for regular sequences $s$ with $\tau_j \geqslant \tau_E$ we get
\begin{eqnarray}
r_q({s}) =Q({\cal N}) / {\tau}_s \;, \qquad 
r_c({s}) = C({\cal N})/ {\tau}_s  \;. 
\label{ratesnomemory}
\end{eqnarray}
\item[{\bf \em c)}]
The generalized memoryless 
configurations~(\ref{mappaENNE30}) 
can be treated in the same way
by replacing the quantities $\tau_s^{\prime}, \tau_s^{\prime \prime}$
of Eqs.~(\ref{unosutauprimo}) and (\ref{unosutausecondo})
with the corresponding average 
first-neighboring {\em group} 
distances and the map $\cal N$ with the
$m$ carriers memoryless map
${\cal M}_s$ of Eq.~(\ref{mappaENNE40}).
In particular, for a generalized memoryless sequences $s$ 
having constant  group lengths $T_g=T_s$ and 
constant group separations $\Delta T_g=\Delta T_s$ 
for all $g$
one easily verifies the following identities
\begin{eqnarray}
r_q({s}) 
&=& {Q({\cal M}_s)}/(T_s + \Delta T_s) 
\;, \label{ratedisblocchi1} \\
r_c({s}) 
&=&{C({\cal M}_s)}/(T_s + \Delta T_s) \;.
 \label{ratedisblocchi2}
\end{eqnarray}
\item[{\bf \em  d)}]
Finally consider the noise attenuation protocols of  Sec.~\ref{s:noiseatt}.
For the sake of simplicity we will focus  on 
the specific example of Fig.~\ref{f:figu5} where the results for
memoryless configuration applies.  In this case the rate is given by
\begin{eqnarray}
r_q({s}) 
&=& {Q(\overline{\cal N})}/{(n \tau +\tau_E)} \;, \nonumber \\
r_c({s}) 
&=& {C(\overline{\cal N})}/{(n \tau + \tau_E)} \;,
 \label{ratedisblocchi}
\end{eqnarray}
with $\overline{\cal N}$ being the map~(\ref{MinputoutputNEW1}) and
with $n \tau + \tau_E$ being the time  intervals which separates  two consecutive 
$A$-carriers.
\end{itemize}

\section{Transmission rate for multiple choice of the
sequence}\label{s:ratechannel}

In this section we analyze the optimal quantum and classical
communication rates $R_{q,c}$ achievable in 
our model  when 
Alice is not restricted to a single given sequence $s$ 
but instead she
has some freedom in selecting the sequence she will use for
the signaling. 

For the sake of simplicity we will assume the set $\cal S$ 
of the allowed sequences 
to be fully characterized by a single
parameter $\tau_{min}$ which bounds the minimum
value for the intervals $\tau_j$ of a sequence $s$ of the set. That is
${\cal S}= {\cal S}(\tau_{min})$  will be the set of all sequences $s$ which satisfy 
$\tau_j \geqslant \tau_{min}$
for all $j$. 
The need of constraining the minimum value of the $\tau_j$
is fundamental if we want our model to have a
non trivial structure (see for instance
Sec.~\ref{s:solvable} and Eq.~(\ref{perfect1}) 
below).
From a more practical  point of view the 
introduction of $\tau_{min}$ follows from the
physical and technological difficulties  in producing sequence
of ordered signals that might arise in realistic 
communication scenarios  (for instance, too close packed
carriers tend to overlap during their propagation,
compromising  the time ordering of the 
sequence).

A natural candidate for $R_{q,c}$ 
is the maximum
of the rates $r_{q,c}(s)$
computed over the sequence $s$ of $\cal S$, i.e.
\begin{eqnarray}
R_{q,c}^{(1)}(\tau_{min}) = \max_{s\in{\cal S}} r_{q,c}({s})
\label{lowerbound}\;.
\end{eqnarray}
A detailed analysis of $R_{q,c}^{(1)}$ is
presented in  Appendix~\ref{s:simplification} 
where
it is shown  how Eq.~(\ref{lowerbound})  
simplifies in the case in which
$\cal S$ contains only regular sequences for which
Eq.~(\ref{ratedis}) applies.
We will see in a moment that for $\tau_{min} \ll \tau_E$ and $\tau_{min} \geqslant \tau_E$ 
the function $R_{q,c}^{(1)}(\tau_{min})$
provides indeed the correct values
of the achievable rates.
For generic $\tau_{min}$ however we
claim that the function $R_{q,c}^{(1)}(\tau_{min})$
does not necessarily tell the whole story about $R_{q,c}$.
On the contrary we propose to compute $R_{q,c}$ as follows
\begin{eqnarray}
R_{q}(\tau_{min}) &=& \lim_{\epsilon \rightarrow 0} \;
\limsup_{T\rightarrow \infty} \;  \max_{s\in{\cal S}}\;\frac{ 
q_{s}(\epsilon,T)}{T}\;,
\label{Formalrate1}\\
R_c(\tau_{min}) &=& \lim_{\epsilon \rightarrow 0} \;
\limsup_{T\rightarrow \infty} \; 
\max_{s\in{\cal S}} \; \frac{c_s({\epsilon},T)}{T}\;,
\label{Formalrate10}
\end{eqnarray}
with $q_s(\epsilon,T)$ and $c_s(\epsilon,T)$ given in 
Eqs.~(\ref{formularate2}) and (\ref{formularate3}).
Equations~(\ref{Formalrate1}) and (\ref{Formalrate10}) 
define proper rates of the communication line
of Sec.~\ref{SEC1} in the sense that, given $\delta >0$ 
and $\epsilon$ arbitrarily small there is allowed sequence ${s}\in {\cal S}$ which, in the limit
of infinite $T$  permit Alice to transfer to Bob at least $(R_q-\delta)T$ qubits
with fidelity $>1-\epsilon$.

Since Eq.~(\ref{lowerbound}) is  obtained from Eqs.~(\ref{Formalrate1}) and 
(\ref{Formalrate10}) 
 by inverting the order of the maximization over ${s}$ with
the limits in $\epsilon$ and $T$ it follows immediately that $R_{q,c}^{(1)}(\tau_{min})$
is a lower bound for
$R_{q,c}(\tau_{min})$ of ${\cal S}$, i.e.
\begin{eqnarray}
R_{q,c}(\tau_{min}) \geqslant R_{q,c}^{(1)}(\tau_{min}) 
\label{lowerbound1}\;.
\end{eqnarray}
An interesting  problem is to understand whether or not the inequality 
in Eq.~(\ref{lowerbound1}) can always be replaced with an identity.
Alternatively one may ask under which conditions on the model parameters 
(i.e. $U_j$, $\cal F$) the transmission rate of ${\cal S}$ can be
computed as the maximum of the rates achievable within a specific choice of~$s$.
In the next section 
we provide a partial answer to these questions
by showing that 
for $\tau_{min} \ll \tau_E$
and $\tau_{min}\geqslant \tau_E$ 
the functions  $R_{q,c}(\tau_{min})$
and $R_{q,c}^{(1)}(\tau_{min})$ coincide.

\subsection{Bounds and asymptotic behavior}\label{s:asymp}

Even without solving the 
maximizations of~(\ref{lowerbound}),~(\ref{Formalrate1}) and~(\ref{Formalrate10}) 
one expects the resulting expressions $R_{q,c}^{(1)}$, $R_{q,c}$
will depend upon
the interplay between  
the relaxation time $\tau_E$  of LE and the characteristic time $\tau_{min}$ 
of  $\cal S$.

A trivial but useful
 upper bound for $R_{q,c}$ follows by observing that
the maximum number $n_s(T)$ of carriers
that can fit in $[0,T[$ cannot be greater than 
$T/\tau_{min}$ and that
$q_s(\epsilon,T), c_s(\epsilon,T)$ 
cannot be greater than
the $\log_2$ of the dimension of ${\cal H}(T)$, i.e.
\begin{eqnarray}
q_s(\epsilon,T), c_s(\epsilon,T)\leqslant n_s(T) \log_2 D\;,
\label{eqnarray}
\end{eqnarray}
with $D$ being the dimension of the Hilbert space of a single carrier.
Replacing the above relations  in Eqs.~(\ref{formalrate1}) and 
(\ref{formalrate10}) gives
\begin{eqnarray}
R_{q,c}(\tau_{min}) \leqslant  
\frac{\log_2 D}{\tau_{min}}\label{upper}
\;,
\end{eqnarray}
for all $\tau_{min}$. 
From Sec.~\ref{s:solvable} it follows that 
this bound is achievable at least if $\cal S$ is such that $\tau_{min}\ll \tau_E$. In this
case in fact the sequence ${s}_0 $ with $\tau_j=\tau_{min}$
for all $j$  allows for carriers that reliably transfer
$\log_2 D$ qubits of information each. Therefore 
from~(\ref{lowerbound}) and 
(\ref{lowerbound1}) we get
\begin{eqnarray}
R_{q,c}(\tau_{min}) = R_{q,c}^{(1)}(\tau_{min})\Big|_{\tau_{min}\ll \tau_E} 
\simeq  \frac{\log_2 D}{\tau_{min}}\;,
\label{perfect1}
\end{eqnarray}
which shows that the rates
 diverge for $\tau_{min}\rightarrow 0$.
An explicit expression  can also be determined 
for $\tau_{min}$ greater than $\tau_E$. 
In fact, according to Sec.~\ref{SEX}, in this
case all the allowed sequences ${s}$ yields the same memoryless mapping
${\cal N}^{\otimes n(T)}$.
Thus the  maximization 
with respect to ${s}$ becomes a simple optimization with respect to the 
average time intervals $\tau_s$ 
and one gets,
\begin{eqnarray}
R_q(\tau_{min})=R_q^{(1)}(\tau_{min})\Big|_{\tau_{min}\geqslant \tau_E} 
&=& \;{ Q({\cal N})}/{\tau_{min}}, 
\label{formalrate1001}\\
R_c(\tau_{min})=R_ c^{(1)}(\tau_{min})\Big|_{\tau_{min}\geqslant \tau_E} 
&=& \;{ C({\cal N})}/{\tau_{min}},
\label{formalrate1002}\end{eqnarray} 
with $Q({\cal N})$ and $C({\cal N})$ the capacities
of Eqs.~(\ref{quantum11}) and~(\ref{cc}),
respectively.

For intermediate value of $\tau_{min}$ 
a lower bound for $R^{(1)}_{q,c}$, and thus for $R_{q,c}$, can be
obtained for instance by focusing on the generalized memoryless
configuration (see Eq.~(\ref{lowerboundmin})) or by considering  the
noise attenuation strategies.  
In this last case it is simpler to consider
only the configurations  described in Fig.~\ref{f:figu5}
and maximizing the rates~(\ref{ratedisblocchi})
with respect to the free parameters  $\tau\geqslant 
\tau_{min}$ and $n\geqslant 1$, e.g.
\begin{eqnarray}
R_{q}^{(1)}(\tau_{min})\geqslant  \;
\sup_{\begin{subarray}{c}
\tau\geqslant \tau_{min}\\
{n\geqslant 1}
\end{subarray}} \;
\frac{Q(\overline{\cal N})}{n\tau + \tau_E} \;,
\nonumber \\
R_{c}^{(1)}(\tau_{min})\geqslant 
  \;
\sup_{\begin{subarray}{c}
\tau\geqslant \tau_{min}\\
{n\geqslant 1}
\end{subarray}} \;
 \frac{C(\overline{\cal N})}{n\tau + \tau_E}
\label{RATEMOD} \;.
\end{eqnarray}

\begin{figure}[t]
\begin{center}
\epsfxsize=.7\hsize\leavevmode\epsffile{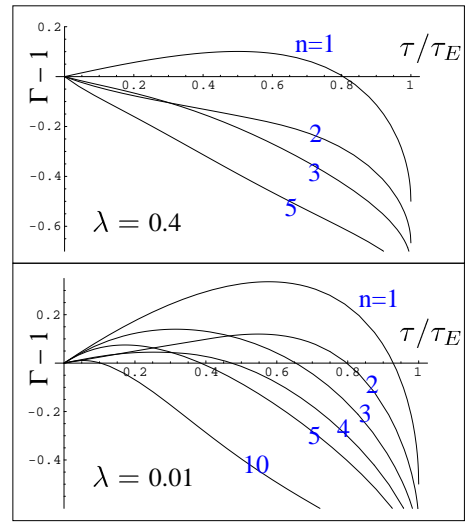}
\end{center}
\caption{Plot of the ratio $\Gamma$ of Eq.~(\ref{RATIO}) as a function of
the dimensionaless parameter $\tau/\tau_E$, 
for different values of the $n$  and for different values
of the environment-carriers coupling constant $\lambda$. 
In the strong coupling regime $\lambda \sim 0$, the attenuation noise protocol
provides a significative improvement of the transmission rate. For instance 
for $\lambda = 0.01$,  $r$ reaches the maximum value of $\sim 1.3$
for $n=1$ and $\tau\sim \tau_E/2$. }
\label{f:fig3}
\end{figure}

\section{An example with qubits}\label{SEC4}

In this section we analyze an example of dynamical model for memory channels where 
both the information carriers C$_j$
and the local environment LE
are qubits.  In this context we will make a comparison between the
noise attenuation protocol of Sec.~\ref{s:noiseatt} and the memoryless configuration.

We will assume the carrier-LE interaction $U_j$ of Eq.~(\ref{seconda3})
to be to a control-unitary
such that 
when the carrier is in $|0\rangle_{C_j}$ 
nothing happens to LE, while when C$_j$ is in $|1\rangle_{C_j}$ 
the environment
undergoes to the 
transformation
\begin{eqnarray}
\Theta(\lambda) \equiv \left( \begin{array}{cc} \sqrt{\lambda} & \sqrt{1-\lambda} \\
\sqrt{1-\lambda} & -  \sqrt{\lambda}
\end{array}
\right) \;,
\label{hadamard}
\end{eqnarray}
with $\lambda\in [0, 1]$ being a parameter which measures the ``intensity''
of the coupling (with low coupling corresponding to $\lambda\sim 1$ and high
coupling corresponding to $\lambda \sim 0$).
Moreover we will assume the relaxation process ${\cal F}=\{ {\cal E}_\tau\}_\tau$ 
acting on LE to be 
described by amplitude damping maps~\cite{CHUANG}  which
takes the state $|1\rangle_{LE}$ to $|0\rangle_{LE}$ with probability $1-\eta(\tau)$
where  $\eta(\tau) \in [0,1]$ is
a non increasing function of $\tau$ with characteristic time
$\tau_E$, i.e.
\begin{eqnarray}
{\cal E}_\tau(|0\rangle_{LE} \langle 0| ) &=&  |0\rangle_{LE} \langle 0| \nonumber \\
{\cal E}_\tau(|1\rangle_{LE} \langle 1| ) &=&  \eta(\tau) \;|1\rangle_{LE} \langle 1|
+(1- \eta(\tau) )\;|0\rangle_{LE} \langle 0| \nonumber \\
{\cal E}_\tau(|0\rangle_{LE} \langle 1| ) &=&  \sqrt{\eta(\tau)} \;|0\rangle_{LE} \langle 1|\;.
\label{DEFNOISE}
\end{eqnarray}
In this example  the stationary state $\sigma_0$ of LE is hence 
$|0\rangle_{LE}$. 
The parameterization of the memory effect 
is given by $\eta(\tau)$,
with $\eta=0$ corresponding to the memoryless case 
(fast environment relaxation) and 
$\eta=1$ corresponding to perfect memory case (no environment relaxation).
In order to have a well defined threshold between memoryless and memory
configuration, in the following we will assume 
\begin{eqnarray}
\eta(\tau) =\left\{ \begin{array}{lll}
1-\tau/\tau_E & &\mbox{for $\tau<  \tau_E$} \\
0 & &\mbox{for $\tau\geqslant  \tau_E$.}
\end{array}
\right.
\end{eqnarray}
Under the above conditions, it 
is possible to show that both the map $\cal N$
of the memoryless case  
and the map $\overline{\cal N}$~(\ref{MinputoutputNEW1}) of the noise attenuation protocol
correspond to a phase damping channel ${\cal P}_g$ where the coherence terms
of the input qubit $\rho$ are degraded by a positive factor $g\leqslant 1$, i.e.~\cite{CHUANG}
\begin{eqnarray}
{\cal P}_g(|\kappa \rangle_{C} \langle \kappa | ) &=&  |\kappa \rangle_{C} \langle \kappa
| \qquad \quad \mbox{for $\kappa =0,1$} \nonumber \\
{\cal P}_g(|0\rangle_{C} \langle 1| ) &=& g \;|0\rangle_{C} \langle 1|\;.
\label{DEFENNE}
\end{eqnarray}
In particular Eq.~(\ref{mappaENNE2}) gives 
${\cal N}={\cal P}_{g_0}$ with 
$g_0=\sqrt{\lambda}$. On the other hand, 
Eq.~(\ref{MinputoutputNEW1}) gives
$\overline{\cal N}={\cal P}_{\overline{g}}$
where $\overline{g}$ is a 
complicated expression~(\ref{gmod}) of 
$\lambda$ and of the parameters $\rho_0$, $n$ and $\tau$ 
(see Appendix~\ref{a:example} for details). 
By appropriately selecting the values of the above quantities
one can make the make  $\overline{\cal N}$ less noisy than 
$\cal N$ by having  $\overline{g}>g_0$. To see if this
corresponds to an increase in the transmission rates $r_{q,c}(s)$ we
can use the results of  Sec.~\ref{s:solvable}.
In the case of the phase damping channels ${\cal P}_g$
the capacities $Q({\cal P}_g)$ and $C({\cal  P}_g)$ of Eqs.~(\ref{quantum11})
and (\ref{cc}) 
can be explicitly computed.
For instance since here the noise
 does not affect the populations associated with 
the computational basis, the classical capacity of the 
phase damping channel~(\ref{DEFENNE})
is optimal  for all values of $g$, i.e. $C({\cal P}_g) = 1$. 
Hence from Eqs.~(\ref{ratesnomemory}) and 
(\ref{ratedisblocchi}) we get
\begin{eqnarray}
r_c(s_0) 
&=& 1 /{\tau_E} \geqslant
{1}/{(n \tau + \tau_E)}= \overline{r}_c \;,
 \label{confronto}
\end{eqnarray}
where $s_0$ is the memoryless sequence with uniform interval $\tau_j=\tau_E$ and
$\overline{r}_c$ is the classical rate of the noise attenuation protocol of
Fig.~\ref{f:figu5}. Equation~(\ref{confronto})
shows that, in the specific 
example considered
here, the  noise attenuation protocol does not improve 
the classical rate of the communication line
with respect to the memoryless case. 
On the other hand the 
quantum capacity of a phase 
damping channel~(\ref{DEFENNE})
is equal to~\cite{DEVETAK}
\begin{eqnarray}
Q({\cal P}_g) = 1-H_2(1/2+g/2)\;,
\label{QUANTUMCAPA}
\end{eqnarray}
where $H_2(x) = -x \log_2 x -(1-x) \log_2 (1-x)$ is the binary entropy function.
In this case, higher values of $g$ corresponds to higher $Q({\cal P}_g)$ and
the rate $\overline{r}_q$ of the noise attenuation protocol can be higher than the rate 
$r_q(s_0)$ of the memoryless case.
To see this 
we studied the ratio 
\begin{eqnarray}
\Gamma = \frac{\overline{r}_q}{r_q(s_0)} = 
\frac{\tau_E}{n\tau + \tau_E}
\frac{ 1-H_2(1/2+\overline{g}/2)}{1-H_2(1/2+g_{0}/2)}
\label{RATIO}\;,\end{eqnarray}
as a function of the variable $\tau/\tau_E$ and for 
for different values of $n$ and $\lambda$. [Here $\overline{g}$ has been optimized with respect
to the no-carrying signal
$\rho_0$].
The results have been plotted in Fig.~\ref{f:fig3} which shows that 
 in the strong  coupling limit $\lambda\sim 0$ one can have  an appreciable increase of $\Gamma$ 
for $\tau\sim \tau_E/2 $ and with $n$ of the order of 5.

\section{Conclusion}\label{s:conclusioni}

We have introduced a communication model 
where memory effects arise from the interaction between the information 
carriers with the channel environment. Different memory effects can be simulated
by varying the time intervals at which the carriers are produced by the sender of the
message. The information rates of the model have been defined and computed in some
extremal cases. 

\appendix
\section{}\label{appendixA}
In this appendix we show how a 
Markovian correlated 
noise~\cite{BOWEN,KRETS,MEMO}
can be derived from the 
mapping~(\ref{APmappaENNE}) by properly
choosing the transformation  ${\cal E}_{\tau_j}$.

Consider the case in which for sufficiently 
big $\tau$  the map ${\cal E}_{\tau}$ 
describes a  decoherent 
process of LE where, given $\{|\ell\rangle_{LE}\}$ an orthonormal basis of ${\cal H}_{LE}$,
one has 
\begin{eqnarray}
{\cal E}_{\tau} (|\ell\rangle_{LE}\langle \ell^\prime| ) =  
\delta_{\ell, \ell^\prime} \; 
|\psi_\ell(\tau)\rangle_{LE}\langle \psi_\ell(\tau)|  \label{DECO}\;,
\end{eqnarray}
with the vectors $\{|\psi_\ell (\tau)\rangle_{LE}\}_\ell$ being not necessarily orthogonal,
and $\delta_{\ell,\ell^\prime}$ being the Kronecker delta.
The condition~(\ref{relax}) can then be satisfied by identifying $\sigma_0$ with
one element of the selected basis (say  $|\ell_0\rangle_{LE}$), and  imposing
$|\psi_\ell(\tau\geqslant \tau_E)\rangle_{LE} = |\ell_0\rangle_{LE}$ for all $\ell$.
In this case the mapping~(\ref{APmappaENNE}) can be expressed in 
terms of the operators 
\begin{eqnarray}
A_{\ell_{1}} &\equiv& {_{LE}\langle} \ell_{1} | U_{1} 
|{\ell_{0}} \rangle_{LE} \\
A_{\ell_{j+1},\ell_{j}} &\equiv& {_{LE}\langle} \ell_{j+1} | U_{j+1} 
|\psi_{\ell_{j}}(\tau_{j})
\rangle_{LE}\;,
\end{eqnarray}
which act, respectively, on the Hilbert space ${\cal H}_{C_{1}}$
and ${\cal H}_{C_{j+1}}$ for $j=1, \cdots, n-1$. They allow us to 
define the probability distribution
\begin{eqnarray}
p^{(1)}_{\ell_1}  &\equiv & \mbox{Tr}_{C_{1}} \left\{ 
A^\dag_{\ell_{1}} A_{\ell_{1}} \right\} 
\label{PROB1}\end{eqnarray}
and the conditional probabilities  
\begin{eqnarray}
p^{(j+1)}_{\ell_{j+1}|\ell_{j}} &\equiv &  \mbox{Tr}_{C_{j+1}} \left\{ 
A^\dag_{\ell_{j+1},\ell_{j}} A_{\ell_{j+1},\ell_{j}} \right\} 
\label{PROB2} \;.
\end{eqnarray}
which satisfies the normalization conditions 
$\sum_{\ell_{j+1}} p^{(j+1)}_{\ell_{j+1}|\ell_{j}} =1$
and $\sum_{\ell_{j}} p^{(j+1)}_{\ell_{j+1}|\ell_{j}} <1$. 
Using these quantities Eq.~(\ref{APmappaENNE}) 
can be finally expressed in compact Markovian form,
\begin{eqnarray}
&\Phi^{(n)}_s(R) = \label{redux1}
\sum_{\ell_1,\cdots, \ell_n} 
p^{(1)}_{\ell_1} \; p^{(2)}_{\ell_2|\ell_1}\;  
\cdots \; \; p^{(n)}_{\ell_{n}|\ell_{n-1}} &\label{MARKOV}
\\
&\times M_{\ell_n,\ell_{n-1}} \nonumber 
\cdots M_{\ell_{2}, \ell_{1}} \;M_{\ell_1}\;  R  \; 
M^\dag_{\ell_1}\; M^\dag_{\ell_{2}, \ell_{1}} \cdots M^\dag_{\ell_n-1,\ell_n} &
\end{eqnarray}
with $M_{\ell_1} \equiv A_{\ell_1} /\sqrt{p^{(1)}_{\ell_1}}$ and
$$M_{\ell_{j+1},\ell_{j}} \equiv A_{\ell_{j+1},\ell_{j}} /
\sqrt{p^{(j+1)}_{\ell_{j+1}|\ell_j}}\;.$$

\section{}
\label{s:simplification}
In this section we analyze
$R_{q,c}^{(1)}$
showing that, if the set $\cal S$ contains only
regular sequences, then the maximization of
Eq.~(\ref{lowerbound})
 can be solved 
by focusing on the 
generalized memoryless configurations.

Consider the subset ${\cal S}_0$ of
 the sequence $s \in {\cal S}$ 
which correspond to the uniform 
generalized memoryless configurations of Sec.~\ref{SEX1} 
characterized by 
constant group distance 
$\Delta T_s = \max\{\tau_{min}, \tau_E\}$.
Since ${\cal S}_0$ is a proper  subset 
of $\cal S$ we have
\begin{eqnarray}
R_{q}^{(1)}(\tau_{min}) &\geqslant&  \max_{s\in{\cal S}_0} r_{q}({s})
\nonumber \\
&=&
\max_{s\in{\cal S}_0}   \frac{Q({\cal M}_s)}{T_s
+ \max\{\tau_{min},\tau_E\}}
\label{lowerboundmin}\;,
\end{eqnarray}
where we used Eq.~(\ref{ratedisblocchi1}) to express $r_q(s)$.
Now, given $s \in {\cal S}$ 
from Eqs.~(\ref{upper1}) and
(\ref{quantum1111})  one gets
\begin{eqnarray}
r_q({s}) &\leqslant&  (1/ {\tau}^\prime_s )
\limsup_{N \rightarrow \infty}  \max_{R} 
\left\{ J (\Phi_{s}^{(N)}, R)/N  
\right\} \nonumber \\
&\leqslant&  (1/ {\tau}^\prime_s )
\limsup_{N \rightarrow \infty} \left\{  \sup_{k\geqslant 1} 
\max_{R^\prime} 
\frac{ J ([\Phi_{s}^{(N)}]^{\otimes k} , R^\prime )}{ k N}  \right\}
\nonumber \\
&=&  (1/ {\tau}^\prime_s )
\limsup_{N \rightarrow \infty} \left\{  \lim_{k\rightarrow \infty} 
\max_{R^\prime} 
\frac{ J ([\Phi_{s}^{(N)}]^{\otimes k} , R^\prime )}{ k N}  \right\}
\nonumber \\
&=&  (1/ {\tau}^\prime_s )
\limsup_{N \rightarrow \infty} \left\{ 
\frac{ Q (\Phi_{s}^{(N)})}{N}  \right\}
\label{NQ3} \;,
\end{eqnarray}
where in the second and in the third line the maximization is performed
over the density matrix $R^\prime$ of $k\times N$ carriers, 
$[\Phi_{s}^{(N)}]^{\otimes k}$ are $k$ copies of the  map
$\Phi_{s}^{(N)}$, and $Q (\Phi_{s}^{(N)})$ is the memoryless quantum 
capacity~(\ref{quantum11})
of the map $\Phi_{s}^{(N)}$.
The second inequality is trivial: it follows from the fact that 
$ \max_R J (\Phi_{s}^{(N)}, R)/N $ can  be obtained
from $\max_{R^\prime} J ([\Phi_{s}^{(N)}]^{\otimes k} , R^\prime )/ (k N) $
for $k=1$.
The identity on the third line instead is a consequence of the fact that 
 $\max_{R^\prime} J ([\Phi_{s}^{(N)}]^{\otimes k} , R^\prime )/ (k N) $
achieves its maximum for $k\rightarrow \infty$.
We can further simplify the above expression  by introducing the time interval
$T_s(N-1) =\sum_{j=1}^{N-1} \tau_j$ 
associated with the first $N-1$ carriers of the sequence $s$
and noticing that
\begin{eqnarray}
\limsup_{N\rightarrow \infty}  \frac{T_s(N-1)}{N} = \tau_s^{\prime\prime} 
\label{NQ1}\;,
\end{eqnarray}
with $\tau_s^{\prime\prime}$ defined as in 
Eq.~(\ref{unosutausecondo}).
Using this result, from Eq.~(\ref{NQ3})  we get
\begin{eqnarray}
r_q(s)&\leqslant & 
\limsup_{N \rightarrow \infty} 
 \frac{T_s(N-1)+\max\{\tau_{min},\tau_E\}}{N \tau_s^\prime}
\nonumber \\
&&\times \limsup_{N \rightarrow \infty} 
\frac{ Q (\Phi_{s}^{(N)})}{T_s(N-1)+\max\{\tau_{min},\tau_E\}} 
\nonumber \\
&\leqslant& \frac{\tau_s^{\prime\prime}}{\tau_s^\prime} \; \sup_{N} 
\frac{ Q (\Phi_{s}^{(N)})}{T_s(N-1)+\max\{\tau_{min},\tau_E\}} \nonumber \\
&\leqslant& \frac{\tau_s^{\prime\prime}}{\tau_s^\prime} \; \sup_{s \in {\cal S}_0} 
\frac{ Q ({\cal M}_s)}{T_s+\max\{\tau_{min},\tau_E\}} \;.
\label{NQ5}
\end{eqnarray}
The ratio $\tau_s^{\prime\prime}/\tau_s^\prime$
is always greater than or equal to one. 
However, if the set $\cal S$ includes only
sequences which are  regular, than for all
$s$ we have $\tau_s^\prime = \tau_s^{\prime\prime}$.
In this case the bounds of 
Eqs.~(\ref{lowerboundmin}) and (\ref{NQ5}) coincides yielding 
\begin{eqnarray}
R_{q}^{(1)}(\tau_{min}) &=& 
 \max_{s\in{\cal S}_0}  \; \frac{Q({\cal M}_s)}{T_s
+ \max\{ \tau_{min}, \tau_E\}} \label{NLB}\;.
\end{eqnarray}
The same derivation applies also for the
classical rate $R_{c}^{(1)}$. In this case one can show
that if $\cal S$ contains only regular sequence then,
\begin{eqnarray}
R_{c}^{(1)}(\tau_{min}) &=& 
 \max_{s\in{\cal S}_0}  \; \frac{C({\cal M}_s)}{T_s
+   \max\{ \tau_{min}, \tau_E\}}\label{NLB1}\;.
\end{eqnarray}

\subsubsection{Asymptotic limit}
It is interesting to note that the above expressions
give the correct asymptotic values of Sec.~\ref{s:asymp}.
For instance for $\tau_{min}\geqslant \tau_E$ we have
${\cal M}_s = {\cal N}^{\otimes m}$ where
$m$ is the number of carriers contained in each
group of the sequence and ${\cal N}$ is the memoryless
map~(\ref{mappaENNE1}). 
Given $s\in{\cal S}_0$ this yields  
\begin{eqnarray}
\frac{Q({\cal M}_s)}{T_s
+ \max\{ \tau_{min}, \tau_E\}} &=&
\frac{m Q({\cal N})}{T_s
+ \tau_{min}} \leqslant \frac{ Q({\cal N})}{\tau_{min}}
\end{eqnarray}
where we used the additivity property
$Q({\cal N}^{\otimes m}) = m Q({\cal N})$
of memoryless channels and the fact that 
group length~(\ref{imp}) is always greater or
equal to $(m-1)\tau_{min}$.
Equation~(\ref{formalrate1001})
finally follows by noticing that 
the rate ${Q({\cal N})}/{\tau_{min}}$
is achieved by the sequence of
${\cal S}_0$ with $\tau_{g,\ell}=\tau_{min}$
for all $g$ and $\ell$.

The limit~(\ref{perfect1}) instead follows by noticing that
the rate ${\log_2 D}/{\tau_{min}}$ can be obtained from
the set ${\cal S}_0$ by using 
$\tau_{g,\ell}=\tau_{min}$ for all $\ell =1, \cdots, m-1$
in the limit of large group, i.e. $m\rightarrow \infty$.
In this case in fact ${\cal M}_s$ is a tensor product of
perfect  memory channels and $T_s = (m-1)\tau_{min}$,
so that 
\begin{eqnarray}
\frac{Q({\cal M}_s)}{T_s
+ \max\{ \tau_{min}, \tau_E\}} &=&
\frac{m \log_2 D }{(m-1)\tau_{min}
+ \tau_{E}}\nonumber \\
&\rightarrow& \frac{\log_2 D}{\tau_{min}}\;.
\end{eqnarray}

\section{}\label{a:example}

To characterize the modified map of $\overline{\cal N}$ we first
solve the system~(\ref{MinputoutputNEW}) by using the following parameterization
for the density matrices element of $\sigma_j$ in the canonical basis
$\{ |0\rangle_{LE}, |1\rangle_{LE}\}$,
\begin{eqnarray}
\sigma_{j} \equiv \left( \begin{array}{cc} 
1-{z_j}  &{x_j} + i {y_j}  \\
{x_j} -i {y_j} & {z_j}
\end{array}
\right) \;,
\label{sigmadec}
\end{eqnarray}
with ${z_j}\in [0,1]$ and ${x_j}$, ${y_j}$ real for all  $j=0,1, \cdots, n$. 
The resulting recursive equation can be simplified by introducing 
the column vectors 
$$\vec{v}_j \equiv (\eta^{-1/4} {z_j}, {x_j})^{T},$$  
$$\vec{w}\equiv (1-p)( \eta^{3/4}(1-\lambda) , \eta^{1/4} \sqrt{\lambda (1-\lambda)})^{T}$$ 
and the $2\times 2$ 
Hermitian matrix 
\begin{eqnarray}
&&A\equiv  (1-p)
\left[ \begin{array}{cc} 
{\eta}( \frac{p}{1-p} -1+2\lambda ) & 
- 2 \eta^{3/4}  \sqrt{\lambda (1-\lambda)}\\
-  2 \eta^{3/4} \sqrt{\lambda (1-\lambda)} &\sqrt{\eta} (\frac{p}{1-p} + 1-2\lambda )
\end{array}
\right], \nonumber
\end{eqnarray} 
where $\eta$ stands for $\eta(\tau)$ and 
$p$ is the population associated with the $|0\rangle_C$ component of the no-carrying
message state $\rho_0$.
In this notation Eq.~(\ref{MinputoutputNEW}) gives the following uncoupled 
equations 
\begin{eqnarray}
y_{j+1} &=& \sqrt{\eta}\; ( 2 p -1 ) \; {y_j}  \label{eq2} \\
\vec{v}_{j+1} &=& A \cdot  \vec{v}_{j} +  \vec{w} \label{eq1} \;,
\end{eqnarray}
 which can be solved analytically. In particular,
imposing the initial condition  $\sigma_0 = |0\rangle_{LE}\langle 0|$ 
(i.e. $x_0=y_0=z_0=0$) the first one gives
${y_j}=0$ for all $j$. The solution of~(\ref{eq1}) instead can be obtained
 in terms of the  eigenvalues $\lambda_{\pm}$ of $A$
and their corresponding eigenvectors $(\alpha_\pm , \beta_\pm)^T$.
Explicitly the eigenvalues of $A$ are 
\begin{eqnarray}
\lambda_\pm &=& \frac{\sqrt{\eta}}{2} [ (1+ \sqrt{\eta}) p \nonumber \\
&&+ (1-p) (1-\sqrt{\eta}) (1-2 \lambda) \pm \Delta ]\;,
\end{eqnarray}
with 
\begin{eqnarray}
\Delta &=& \{ 4 \sqrt{\eta} (1-2p) \\ &&+ 
[ (1+ \sqrt{\eta}) p  + (1-p) (1-\sqrt{\eta})(1-2\lambda)]^2 \}^{1/2}
\nonumber \;.
\end{eqnarray}
The corresponding eigenvectors $(\alpha_\pm, \beta_\pm)$ have instead the following
components 
\begin{eqnarray}
\alpha_\pm &=& \eta^{1/4} (1-p) \sqrt{\lambda (1-\lambda)}/N_\pm\;,
\label{eigenv} \\
\beta_\pm &=& [(\sqrt{\eta}-1) p -(1-p) (1-2\lambda)(1+\sqrt{\eta}) \mp \Delta]/N_\pm \;,
\nonumber
\end{eqnarray} 
with the normalization coefficient 
\begin{eqnarray}
&N_\pm =  \{ 16 (1-p)^2 \sqrt{\eta} \lambda (1-\lambda) &
\label{normalizzazione}\\
&+ [ (\sqrt{\eta}-1)p -(1-p) (1-2\lambda) (1+\sqrt{\lambda}) \mp \Delta]^2 \}^{1/2}&
. \nonumber
\end{eqnarray}
In particular, for $|\lambda_\pm|< 1$ one has~\cite{NOTA11}
\begin{eqnarray}
\vec{v}_{j} &=& A^j \cdot \vec{v}_{0} +  \sum_{k=0}^{j-1} A^k \cdot \vec{w} 
 = \frac{\openone- A^j}{\openone-A} \cdot \vec{w} 
\label{AEQ1}
\end{eqnarray}
and thus
\begin{eqnarray}
{z_j} &=& \eta^{3/4} (1-p) \;\left[ \eta^{1/4} (1-\lambda) 
u^{(j)} + \sqrt{\lambda(1-\lambda)} t^{(j)}\right] \nonumber \\
x_j &=& \eta^{1/2} (1-p) \;\left[ \eta^{1/4} (1-\lambda) 
t^{(j)} + \sqrt{\lambda (1-\lambda) }v^{(j)}\right] \nonumber \;,
\end{eqnarray}
where
$u^{(j)} = \xi^{(j)}_+ \alpha_+^2 +  \xi^{(j)}_- \alpha_-^2$, 
$v^{(j)} = \xi^{(j)}_+ \beta_+^2 +  \xi^{(j)}_- \beta_-^2$,
and $t^{(j)} = \xi^{(j)}_+ \alpha_+\beta_+ +  \xi^{(j)}_- \alpha_-\beta_-$
with 
\begin{eqnarray}
\xi^{(j)}_\pm = \frac{1-(\lambda_\pm)^j}{1-\lambda_\pm} 
\label{xi}\;.
\end{eqnarray} 
Setting $j=n$ and replacing the above expressions into (\ref{sigmadec}) 
we obtain the modified  state of LE $\sigma_n$ after $n$ successive interactions with
$\rho_0$. Using the definition (\ref{MinputoutputNEW1}) one verifies that 
$\overline{\cal N}$ is a phase damping channel~(\ref{DEFENNE})  characterized by a 
damping factor
\begin{eqnarray}
\overline{g}= \sqrt{\lambda} -2 \; ( \sqrt{\lambda} \;z_n - \sqrt{1-\lambda} \; x_n )\;.
\label{gmod}
\end{eqnarray}

\acknowledgments

I would like to thank Rosario Fazio for  remarks and suggestions:
without his encouragement this work would never been completed.
Moreover I would like to thank Chiara Macchiavello and 
 Massimo Palma for their comments and discussions.
In particular, thank to Massimo for pointing out Refs.~\cite{SCARANI}.


\begin{thebibliography}{99}
\bibitem{NOTAZERO}The integer $n_s(T)$ is uniquely defined
by the inequalities
\begin{eqnarray}
\sum_{j=1}^{n_s(T)-1} \tau_j \leqslant T < \sum_{j=1}^{n_s(T)} \tau_j
\label{defenne}\;.
\end{eqnarray}
\bibitem{NOTASUP}
The presence of
$\limsup_{T \rightarrow \infty}f(T) 
= \lim_{T\rightarrow 
\infty} \sup_{t\geqslant T} f(t)$
in place of a regular limit over $T$ in
 Eq.~(\ref{formalrate1})
stems from the fact that
one is considering the
maximum over all possible
rate achievable. 
Analogous definitions
applies in the case
of channel capacities
of memoryless channels (see
for instance 
Refs.~\cite{KRETS,KRETS1,BEN,BARNUM}).
\bibitem{NOTALIM} The inequalities
(\ref{upper1}) and (\ref{lower1}) 
do not apply, respectively, in the pathological situation where
$(\tau^\prime_s, Q_s) =(0,0)$ and 
$(\tau^{\prime\prime}_s, Q_s)=(0,0)$. 
These cases are hence excluded from the  analysis.
\bibitem{NOTA11} The case in which one or both the eigenvalues of $A$ have unitary
modulus can be treated analogously.
\end{thebibliography}
\end{document}